\documentclass[runningheads]{llncs}
\usepackage{graphicx}
\usepackage{csquotes}
\usepackage{enumitem}

\usepackage{amsmath} % unverzichtbare Mathe-Befehle
\usepackage{amssymb} % viele Mathe-Symbole
\usepackage{mathtools} % Erweiterungen für amsmath

\usepackage{caption}
\usepackage{subcaption}
\usepackage{todonotes}
\usepackage{cite}
\usepackage[group-minimum-digits=4]{siunitx}

\DeclareMathOperator{\avg}{\text{tm}}

\DeclareMathOperator{\selfexp}{\text{SelfExp}}
\DeclareMathOperator{\score}{\text{score}}
\DeclareMathOperator{\view}{\text{view}}
\DeclareMathOperator{\datapoint}{p}
\DeclareMathOperator{\projpoint}{\datapoint^\prime}
\DeclareMathOperator{\proj}{\text{proj}}
\DeclareMathOperator{\projset}{\text{proj\_set}}

\begin{document}
\title{
% Simulation study on and improvement of outlier detection methods in data cubes
RODD: Robust Outlier Detection\\in Data Cubes
% \thanks{\textcolor{red}{Affiliation: This work has been partly supported by the Research Center Trustworthy Data Science and Security (https://rc-trust.ai), one of the Research Alliance centers within the https://uaruhr.de.}}
}

% \thanks{Supported by organization x.}

\author{Lara Kuhlmann$^*$\inst{2,3}\orcidID{0009-0006-2934-0203} \and
Daniel Wilmes$^*$\inst{1}\orcidID{0009-0000-6358-7732} \and
Emmanuel Müller\inst{1,4}\orcidID{0000-0002-5409-6875}  \and
Markus Pauly\inst{2,4}\orcidID{0000-0002-0976-7190} \and
Daniel Horn\inst{2,4}\orcidID{0000-0002-5208-0482}}
\titlerunning{RODD: Robust Outlier Detection in Data Cubes}
\authorrunning{Kuhlmann, Wilmes, Müller, Pauly \& Horn}
\institute{
Department of Computer Science, TU Dortmund University,
Dortmund, Germany, \email{\{daniel.wilmes, emmanuel.mueller\}@cs.tu-dortmund.de} \and
Department of Statistics, TU Dortmund University,
Dortmund, Germany, \email{lara.kuhlmann@tu-dortmund.de}, \email{\{pauly, dhorn\}@statistik.tu-dortmund.de} \and
Graduate School of Logistics, Department of Mechanical Engineering, TU Dortmund University, Dortmund, Germany \and 
Research Center Trustworthy Data Science and Security, TU Dortmund University,
Dortmund, Germany
}
\maketitle              % typeset the header of the contribution
\def\thefootnote{*}\footnotetext{These authors contributed equally to this work}\def\thefootnote{\arabic{footnote}}

%
%\vspace{-0.5cm}
\begin{abstract}
Data cubes are multidimensional databases, often built from several separate databases, that serve as flexible basis for data analysis. %, where the dimensions allow to specify hierarchical structures.
Surprisingly, outlier detection on data cubes has not yet been treated extensively. %We study outlier detection in data cubes. %can provide valuable insights into data and thus, is an important task in data analysis.  
%A special, but yet important setting is the detection of outliers. 
%Interestingly, outlier detection on data cubes is not widely studied.
In this work, we provide the first framework to evaluate robust outlier detection methods in data cubes (RODD). We introduce a novel random forest-based outlier detection approach (RODD-RF) and compare it with more traditional %outlier detection 
methods based on robust location estimators. We propose a general type of test data and examine all methods %on data cubes 
in a simulation study. 
Moreover, we apply ROOD-RF to real world data.
The results show that RODD-RF can lead to improved outlier detection. %$yields the highest AUC score.


\end{abstract}
\section{Introduction}
\label{sec:introduction}

The amount of data created, captured, copied, and consumed worldwide annually has increased rapidly over the past years: from 2 zettabytes in 2010 to around 79 zettabytes in 2021 \cite{holst2021volume, pauleen2017does}. Both in public data and in internal data of companies, the detection of rare events in form of outliers can provide valuable insights into customer opinions, market developments and processes \cite{11420_13905}. 
% Outliers (depending on the context also called anomalies) are data points that significantly differ from other data points and thus, hint at a change in the data-generating process. For business analysts, outliers in real-world datasets often indicate either a problem or an opportunity and thus, can provide valuable insights 
% \todo{"valuable insights" steht schon 5 Zeilen darüber}
% when correctly identified.  
Thereby, the data is often not available in a single, but in several databases and the data is merged in a separate decision support database called \emph{data warehouse}. In contrast to a traditional database management system, which serves to record transactions, a data warehouse allows the extraction of knowledge from the data through analysis. A data warehouse is commonly modeled via a so-called \emph{data cube} \cite{gray1997data}. 

A data cube allows us to view the data in multiple dimensions.
\begin{figure}
    % \label{fig:datacube}
    \centering
    \includegraphics[width=\textwidth, trim= 0 25mm 0 0]{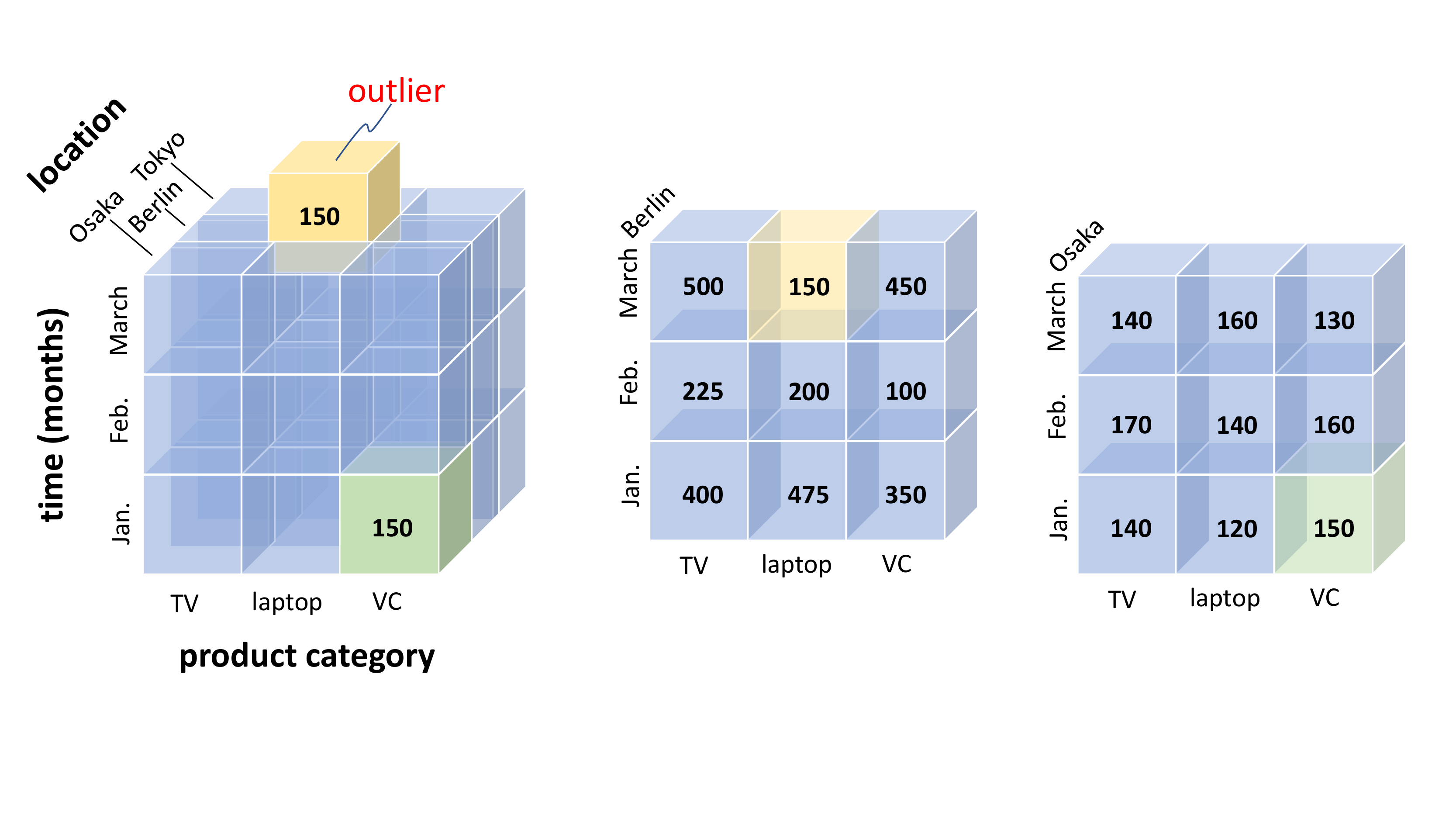}
    \caption{Left part: Example of a data cube; middle part: a slice of the data cube displaying the sales in Berlin; right part: a slice of the data cube displaying the sales in Osaka.}
     \label{fig:Data Cube}
\end{figure}
An example of a data cube is given in Figure~\ref{fig:Data Cube}. Consider the sales data 
%(numerical value) 
of an electronics company viewed in the categorical dimensions of product category, month and city. We construct the following toy example: The type of product can either be a laptop, a vacuum cleaner or a TV; a month can be (for sake of clarity) January, February  or March of 2022; the considered cities are Osaka, Tokyo and Berlin. The values of the categorical dimensions define a cell of a data cube in which a numerical value $y$, in our case the concrete sales number, is stored: e.g. the cell marked in green, names the sales number of all vacuum cleaners in March 2022 in Osaka, which is in our example $150$. 
In this example, another cell (laptop, March, Berlin; marked in yellow) has the same sales number $150$. How can we determine whether one of these data points is an outlier or not? What should methods for robust outlier detection in data cubes (RODD) look like?

We do not have a ground truth of outliers and thus, only methods of unsupervised learning are applicable. 
One data-driven way is to compare each point with 'related' points. In the middle part of Figure \ref{fig:Data Cube}, a slice of the data cube shows the sales in Berlin for the different months and products and in particular contains the marked yellow cell. Although the sales number 150 is not the lowest value in the slice, it is considerably lower than the numerical values in the column and row of the considered cell: The sales of TVs and vacuum cleaners in March and also, the sales of laptops in February and January are higher. Thus, the sales of laptops in Berlin for March would be expected to be a lot higher than 150 and it is reasonable to view this cell as an outlier. % especially, if the same pattern applies to other slices/views which contain the yellow cell (regarding laptop and March).
Note that for the green cell (right-hand side of Figure~\ref{fig:Data Cube}), the view of the data cube regarding Osaka does not show any of the peculiarities mentioned above: the cell's value does not deviate much from the other values of the view. %in the sense described above. 
Thus, the value $y=150$ is anomalous only for the cell marked in yellow, not for the one marked in green. Because of this, statistical methods like simple extreme value analysis \cite{walfish2006review} fail in this setting: 150 is no extreme value. Clustering the values in the slice regarding the city Berlin (middle part of Figure~\ref{fig:Data Cube}) into the clusters  $\{100,150,200,225   \}$ and $\{350, 400, 450, 475, 500  \}$, 
would rather yield that $y=150$ is contained in a cluster. Hence, also techniques such as distance-based outlier detection \cite{knorr1997unified} and density-based outlier detection like 
local outlier factor (LOF) \cite{breunig2000lof} fail. This poses a challenge to the detection of outliers in data cubes.

% traditional outlier mining technique such as distance-based outlier detection \cite{knorr1997unified} and density-based outlier detection like LOF (local outlier factor)  \cite{breunig2000lof} fail
% This is why traditional outlier mining techniques such as distance-based outlier detection \cite{knorr1997unified} and density-based outlier detection like LOF (local outlier factor)  \cite{breunig2000lof} fail. Also statistical methods like simple extreme value analysis \cite{walfish2006review} fail in this setting: 150 is no extreme value. This poses a challenge to the detection of outliers in data cubes.

Another challenge is that many algorithms for outlier detection are only applicable in the presence of numerical values \cite{wang2019progress}. In our  data cubes scenario, however, there are categorical values that determine the position of a cell in the data cube and there is only one numerical value, in our case a sales value. 
%which is stored in the data cube.
%Therefore, an algorithm such as LOF is not applicable in this scenario.

Given these observations, an intuitive way for RODD is the following: Use any sensible prediction method to estimate each cell's sales number. If the estimated value $\hat{y}$ for a cell differs too much from the actual value $y$, mark it as outlier. 
% In comparison, the sale of 100 vacuum cleaners in January is also lower than the other cells in its column and row, but the difference to the values is less. That makes it conspicuous, but maybe not enough to be labeled as an outlier.
Based on this intuition, 
\cite{sarawagi1998discovery} resembles an ANOVA approach 
\cite{st1989analysis}
for RODD. 
Using trimmed mean estimates $\hat{y}$,
%the computation of an estimated sales value $\hat{y}$ which is compared with the actual sales value $y$. T
the deviation between the actual and the estimated sales value, $r \coloneqq \lvert y - \hat{y} \rvert$, is normalized by the spread $\sigma$ of the data. This yields the so-called  \emph{SelfExp} value which defines the outlierness of an event: $\selfexp(y) \coloneqq \dfrac{r}{\sigma}$. All SelfExp values that exceed a certain threshold $\tau$ are considered as an outlier. Whereas \cite{sarawagi1998discovery} only considers the usage of a specific trimmed mean for the computation of $\hat{y}$, the approach actually works for other sensible estimators as well. In particular, we propose to use a random forest (RF) regressor \cite{breiman2001random} for the computation of $\hat{y}$ as it was proven to be a robust method for many regression tasks, see, e.g.,  \cite{el2022random,huang2020travel,cootes2012robust}. We call this method RODD-RF. 

The paper is structured as follows: Section 2 presents related work. Section 3 introduces a framework for RODD. Thereafter, we discuss different approaches for estimating the value for a data cube cell (Section 4). A simulation study evaluates the performance of the different estimators (Section 5). The results of the study and its limitations are discussed (Section 6) and the RODD-RF method is validated on a real-world dataset (Section 7). Finally, a conclusion and an outlook are provided (Section 8).

\section{Related Work}

% In this section, we review different outlier mining paradigms and explain how our RF-based approach relates to them.

\emph{Supervised} outlier detection methods require the labeling of anomalous training data. Examples are given by \cite{vargaftik2021rade, mohandoss2021outlier} who train a RF classifier on labeled outliers. In contrast, \emph{unsupervised} methods do not require the labeling of anomalies. RF-based approaches in an unsupervised scheme are proposed in \cite{mensi2022using, zhang2006anomaly, liu2008isolation, liu2012isolation}. \emph{Model-based} outlier detection methods learn parameters of a model based on the given data. Here one makes the assumption that the data were indeed generated by the used model. A subcategory of this method is a Gaussian mixture model \cite{yang2009outlier} whose parameters are learned via Expectation maximization \cite{dempster1977maximum}. A second subcategory are regression-based methods. Here the prediction of the regressor is compared with the actual value of the data point \cite{park2015regression, wang2019progress}. 
A drawback of model-based outlier detection methods is the fitting of the model parameters to an assumption made about the data. This assumption may not necessarily hold. For this reason, \emph{non-parametric} outlier detection methods were developed. Here no parameters are fitted to a model. Among them are distance-based outlier detection \cite{knorr1997unified}, the Local Outlier Factor (LOF) \cite{breunig2000lof, jin2006ranking} and Kernel density estimation methods (KDE) \cite{pavlidou2014kernel, latecki2007outlier}. 

A third paradigm for outlier detection is the usage of \emph{neural networks}. The problem can be considered as a supervised classification task where outliers have their own labels \cite{wang2019progress}. However, as labels are mostly not available, many methods focus on unsupervised learning \cite{ruff2018deep, chalapathy2018anomaly, zong2018deep}.
One option is an autoencoder approach \cite{zhou2017anomaly, hawkins2002outlier, chen2017outlier, andrews2016detecting}.
To counter the problem of missing labels, semi-supervised approaches which utilize the information of labeled data points have been proposed  \cite{ruff2019deep, oliver2018realistic}.

%Our proposed method of using a RF regressor to predict the estimated numerical value $\hat{y}$ falls under the category of supervised, model-based outlier detection. Remind that many of the stated methods cannot be applied for outlier detection on a data cube due to the already stated challenge of a mixture of categorical and numerical values of a data point.
In the case of data cubes applied in an industrial context, most of the common outlier detection methods cannot be applied. Companies usually have no information about outliers in their data prior to applying an outlier detection method. Thus, all supervised or semi-supervised methods can be disregarded. Moreover, distance-based methods could be applied, but they would neglect the categorical data and subsequently valuable information. Therefore, we focus on unsupervised outlier detection methods, specifically those that calculate an expected value for a data cube cell and compare it with the actual value. Subsequently, our research subject is the comparison and performance evaluation of estimators, such as our proposed RF regressor.

\section{A Framework for Outlier Detection in Data Cubes}
\label{sec:framework}
% \todo{vielleicht das Kapitel so aufbauen: Durch den besonderen Aufbau von Data Cubes sind kaum Standard Methoden für O.D. anwendbar; dann Aufbau von Datacubes mit verschiedenen Views erklären; daher wird auf das Residuum von y - yDach zurückgegriffen;
% vielleicht das Wort SelfExp noch gar nicht aufgreifen, könnte in Kaptiel 4 als eine Methode für die Residuums Berechnung präsentiert werden}

\begin{figure}
    \centering
    \includegraphics[width=\textwidth]{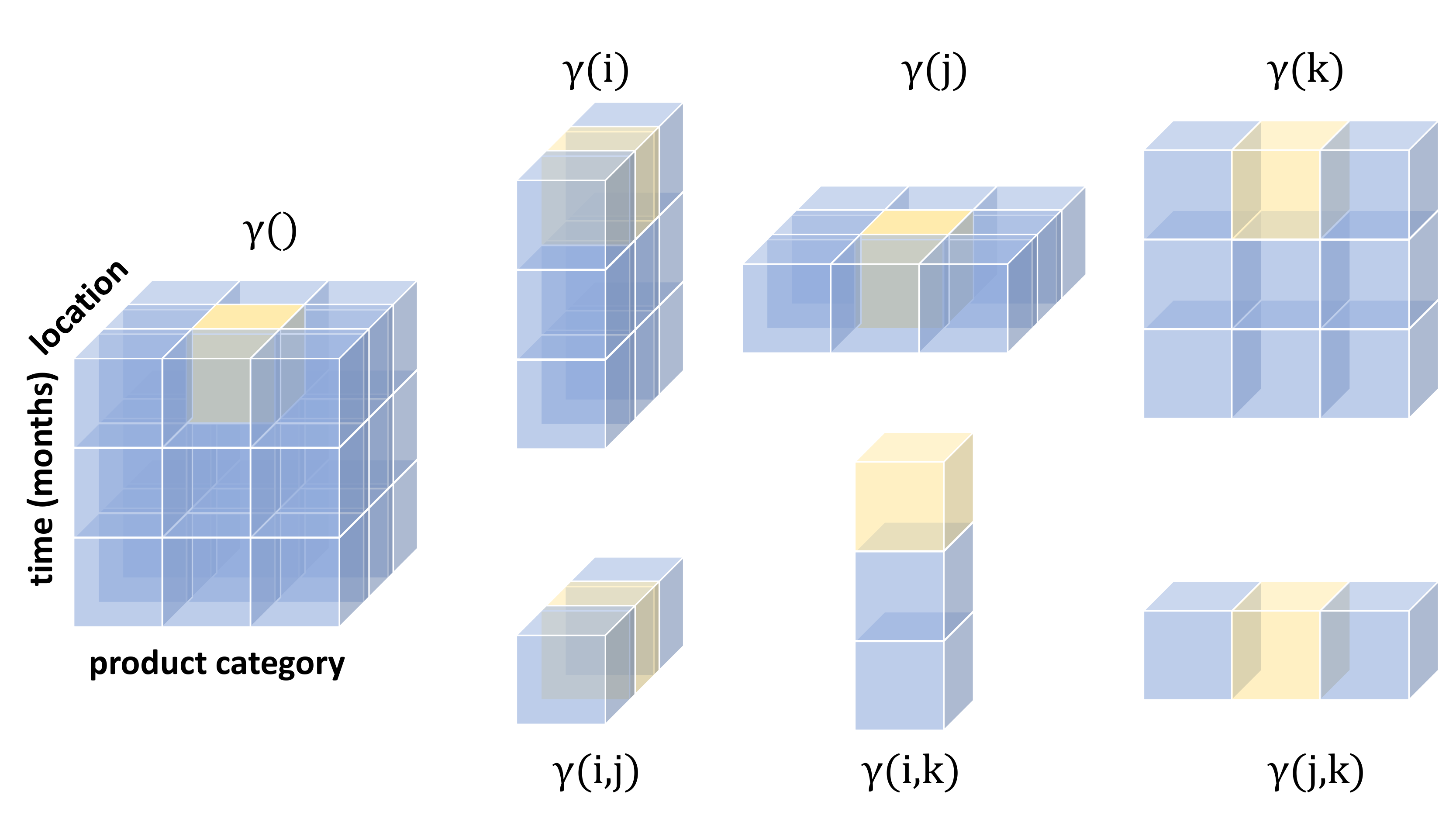}
    \caption{Example: Different views on a data cube. Each of the seven views on the data cube displays the relevant cells which are used to compute a corresponding model coefficient $\gamma$. All seven model coefficients are then needed to determine the estimated value $\hat{y}$ of the yellow cell.} 
    \label{fig:views}
\end{figure}

%SelfExp is a method to detect outliers in data cubes. 
A data cube is an $n$-dimensional database DB. Let $D \coloneqq \{d_1, \dots, d_n\}$ be the set of all dimensions where each dimension is categorical. Each data point $\datapoint \in DB$ gets assigned a numerical value $y \coloneqq f(\datapoint)$ which is stored in the data cube. For example, $f(\datapoint)$ can be a sales number. 
%As stated in Section~\ref{sec:introduction}, the search for outliers in data cubes is a challenging task: due to the presence of both categorical and numerical values at the same time, traditional outlier mining techniques are not applicable. Furthermore, 
%as demonstrated in the toy example of section~\ref{sec:introduction}, 
%the definition of an outlier cannot rely on any clustering or traditional statistically related paradigms.
In the following, we abstract a framework for outlier detection in data cubes from the technique presented in \cite{sarawagi1998discovery}: In a high-level view, one computes for each numerical value $y$ a corresponding estimated value $\hat{y}$ and also the spread of the data. This is incorporated into an outlierness score  $\score(y) = {\lvert y - \hat{y}  \rvert}/{\sigma}.$ 
If $\score(y)$ exceeds a threshold $\tau$, the value $y$ and its corresponding data point $\datapoint$ is considered an outlier. Below we present a general way for the determination of $\hat{y}$. %The computation of $\sigma$ is described in Section~\ref{sec:estimators}.
To this end, let us introduce some notation: Given a data point $\datapoint \in DB$, the projection of $\datapoint$ to the dimensions $D^\prime \subsetneq D$ is noted as $\projpoint \coloneqq \proj(\datapoint, D^\prime)$. We now resemble a general analysis of variance (ANOVA) approach \cite{st1989analysis} as follows: To describe the effect of the subset of dimensions $D^\prime$ on the sales number $y(p)$, we 
assign a model coefficient $\gamma(\projpoint)$ to each projection.  Here, a typical choice for the model coefficients $\gamma(\projpoint)$ is the (trimmed) mean over all values in the projection $\projpoint$, see the illustrative example below. As sales number are often of multiplicative nature \cite{sarawagi1998discovery}, we consider its logarithmic value: We thus estimate the 
logarithmic sales number for $\datapoint$ by 
summing the effects $\gamma(\projpoint)$ for every projection, i.e via
\begin{equation} 
    % \label{eqn:highlevel}
    \log \hat{y}(\datapoint) = \sum_{\projpoint \in \projset(\datapoint)} \gamma(\projpoint),
\end{equation}
where the sum runs through all possible projections $\projpoint$ of $\datapoint$. %, consider the following general formula:This means, for every projection of a point $\datapoint$ there is a corresponding model coefficient $\gamma(\projpoint)$. The sum of the model coefficients yields $\log \hat{y}(\datapoint)$. This approach resembles an analysis of variances (ANOVA) approach \cite{st1989analysis}.
Each model coefficient takes into account a subset $D^\prime \subsetneq D$ of dimensions and a corresponding selection of data cells. A model coefficient can be viewed as an estimated value over the data cells which are picked by it. Let us show this via an example: Assume in Figure~\ref{fig:views} we have the same setting as in our toy example of Section~\ref{sec:introduction} and want to compute the estimated numerical value (sales value) $\hat{y}$ of the data point $\datapoint$=(laptop, March, Berlin), marked yellow, which we abbreviate $\datapoint=(i,j,k)$. Each model coefficient now computes an estimated value over a \emph{view}, which means a subset of dimensions and data points of the data cube. 

For our tuple $\datapoint= (i,j,k)$ the set $\projset$ of all projections of $\datapoint$ is given by $\projset(\datapoint) = \{ (i,j), (i,k), (j,k), (i), (j), (k), () \}$. Note that the empty projection $()$ is allowed. We then collect all model coefficients in the set $\{ \gamma(\projpoint) \mid \projpoint \in \projset(\datapoint)  \} = \{ \gamma(i,j), \gamma(i,k), \gamma(j,k), \gamma(i), \gamma(j), \gamma(k), \gamma() \}$.
Now that we know which model coefficients are to be computed, we have to determine for every model coefficient the corresponding set of data points on which it is computed.
Given a subset of dimensions $D^\prime$, the view of a model coefficient $\gamma(\projpoint)$ is then defined as  $\view(\gamma(\projpoint)) \coloneqq \{ \datapoint \in DB \mid \projpoint  \text{is projection of} \, \datapoint  \}$ (cf. Figure~\ref{fig:views} with data point $\datapoint=(i,j,k)$). % This is exactly what we see in Figure~\ref{fig:views}, where we consider the data point $\datapoint=(i,j,k)$.
We now compute $\log \hat{y}(p)$, and by this the deviation $\lvert y(\datapoint) - \hat{y}(\datapoint)  \rvert$ by finding all sufficient model coefficients and then computing every model coefficient on the corresponding set of data points. E.g., for $\gamma()$ we consider the full toy data cube as the empty tuple (). It is a projection of every data point and thus resembles a global effect. Finally, we compute:
\begin{equation}
    \log \hat{y}(\datapoint) = \gamma(i,j)+ \gamma(i,k)+ \gamma(j,k)+ \gamma(i)+ \gamma(j)+ \gamma(k)+ \gamma() .
\end{equation}

% $y(v)$ is compared to its estimated value $\hat{y}(v)$ (in our application a sales number). This so called \emph{residual} $r(v) \coloneqq |y(v) - \hat{y}(v)|$ is normalized by the spread of data given $v$. 
To obtain a standardized outlier score, we divide the resulting absolute residual $\lvert y(\datapoint) - \hat{y}(\datapoint)  \rvert$ 
by a suitable measure of dispersion $\sigma = \sigma(\datapoint)$, see Section~\ref{sec:estimators} for a specific choice. %The way $\hat{y}(v)$ and $\sigma(v)$ are calculated is described below. 
% Whether a specific (sales) numerical value is viewed as an outlier, can now be determined via the so called \emph{SelfExp} value:
The comparison with a treshhold $\tau$ leads to the outlierness score
\begin{equation} \label{eqn:selfexp}
    \selfexp(y(\datapoint)) = \max \left(
    \dfrac{ |y(\datapoint) - \hat{y}(\datapoint)| }
    {\sigma(\datapoint)} - \tau, 0
    \right).
\end{equation}
Though our framework is more general we have retained the notion SelfExp from the original paper \cite{sarawagi1998discovery}. %Here, 'Exp' stands for 'expected' as the method uses estimators for the expected (sales) value. 
If the normalized residual exceeds a certain threshold $\tau$, i.e. if 
$\selfexp(y(\datapoint))>\tau$, 
a data point $\datapoint$ is considered as outlier. Usually, $\tau$ is set to 2.5. This refers to the fact, that in the case of a normal distribution almost $99\, \%$ %(more concrete $98.8 \, \%$)
of the data are within the $2.5 \,$ fold standard deviation around the mean.

\section{Robust Estimation of Data Cube Cells}
\label{sec:estimators}

%In this section, we give concrete instantiations of possible estimators $\hat{y}$ 
%for our RODD framework. This includes the SelfExp estimator \cite{sarawagi1998discovery}, other trimmed mean-based estimators, a median-based estimator and our proposed RODD-RF estimator. Moreover, we explain our calculation of $\sigma$. 

%As described in the previous section, outlier detection in data cubes mostly relies on the standardised residuum of $\lvert y(\datapoint)$ - $\hat{y}(\datapoint) \rvert$. To explore different implementations of this equation, the focus had to be on either the calculation of the expected value $\hat{y}$ or of  the standard deviation. Alternatives for the standard deviation were considered, but not looked into deeper. 
%On the one hand, the standard deviation was obliged to never being zero. On the other hand, altering the standardization measure would have led to a loss of comparability of the outlier detection methods.
%Thus, this section presents different estimators $\hat{y}$.

Building on the equations introduced in Section~\ref{sec:estimators}, we can now define different outlier detection methods by giving concrete estimators for $\hat{y}$ and $\sigma$. In the case of SelfExp \cite{sarawagi1998discovery}, the computation of the model coefficients is done by 
averaging. Starting with the general effect represented by $\gamma()$ we compute %computing an   \emph{expected} value for each model . Following our example :
%\begin{equation*}
%    %\gamma \coloneqq 
%    \gamma() = 
%    \underset{_{\datapoint}}{\avg}\log %y(\datapoint)
%    \label{equation: trimmed mean}
%\end{equation*}
%as 
the $12.5 \, \%$-trimmed  mean ($\avg$) of the logarithm of all values $y(\datapoint)$ in the data cube. The $\avg$ is used for sake of robustness and neglects the $12.5 \, \%$ largest and smallest values. Going back to our example with $\datapoint = (i,j,k)$ the model coefficients for the projections on one dimension can be computed as  %In case of  $\gamma(i)$ we set
\begin{equation*}
    \gamma(i) = \underset{_{\projpoint = (j,k) }}{\avg} \log y(\datapoint) - \gamma(),
\end{equation*}
% Here $v^\prime=(j,k)=(\text{Osaka, January 2019})$ and $y(t^\prime)$ is correspondingly the sales number of all sold products in Osaka in January
% 2019. In other words, $y(v^\prime)$ aggregates all sales values along dimension $A$, i.e. along dimension \enquote{product}.
and for a 2-dimensional coefficient $\gamma(i,j)$ as
\begin{equation*}
    \gamma(i,j) = \underset{_{\projpoint = (k) }}{\avg} \log y(\datapoint) - \gamma(i)- \gamma(j)- \gamma().
\end{equation*}
All other coefficients are computed in a similar fashion, e.g. coefficient $\gamma(i,k)$ is obtained by averaging over the logarithm of all aggregated sales data of the dimension time and then subtracting the global coefficient $\gamma()$ and all coefficients that depend on $i$, respectively $k$. As already mentioned, the way of computing $\hat{y}(\datapoint)$ resembles an ANOVA, more concretely for this example a three-fold ANOVA with twofold interaction effects \cite{searle2016linear}. 
The same construction pattern is applied to higher dimensional coefficients.

%The only missing component for computing the value $\selfexp(y(\datapoint))$ is $\sigma(\datapoint)$.
Following \cite{sarawagi1998discovery}, we define $\sigma(\datapoint)$ via
\begin{equation}
    \sigma(\datapoint)^2 = (\hat{y}(\datapoint))^\rho.
\end{equation}
Here, $\rho$ is chosen using the maximum likelihood principle under the assumption that the data are normally distributed with a mean value $\hat{y}(\datapoint)$. From this we can derive that $\rho$ has to satisfy the following equation \cite{sarawagi1998discovery}:
\begin{equation} \label{eqn:variance}
    \sum_{\datapoint } \left(\dfrac{(y(\datapoint)-\hat{y}(\datapoint))^2}{(\hat{y}(\datapoint))^\rho} \cdot \log(\hat{y}(\datapoint)) - \log(\hat{y}(\datapoint)) \right) = 0,
\end{equation}

To our knowledge, the SelfExp estimator was the only instantiation of robust outlier detection in data cubes 
so far. But it has never been systematically evaluated nor challenged. The trimmed mean as an estimator is intuitive 
%and through the trimming robust against outliers, 
but there might be more effective ways to calculate a robust estimator. We discuss further approaches for RODD in the sequel. To distinguish the SelfExp approach from the others, we introduce the notation $\hat{y}\textsubscript{S75}$ for the 
SelfExp (sales value) estimator, as it neglects $25 \%$ observations when computing the trimmed means. 

%After having introduced the SelfExp estimator $\hat{y}\textsubscript{S}$, the following paragraphs will present alternative approaches for deriving $\hat{y}$.  
%\todo{why do we need alternatives? What are the disadvantages of the original estimator?}

%Since the publication of the SelfExp method in 1998,

Since the SelfExp publication in 1998 more advanced machine learning regression methods were developed. Machine learning models were reported to often outperform linear regression methods (that are underlying an ANOVA) and gained popularity among both researchers and practitioners \cite{ardabili2020advances}. One important novelty was the introduction of the RF \cite{friedman2006recent}. The RF by Breiman \cite{breiman2001random}
is an ensemble regressor that applies a bagging approach. It consists of multiple decision trees, which are trained on samples from the data set. 
Each tree predicts a value and in the last step, an average across all these values is calculated and serves as the final prediction.
%\textit{There are several model parameters, such as the number of features that were used as input for a tree and the depth of a tree. Tuning these parameters can improve the model's performance, but it was often noticed by researchers that it comes with high computational cost \cite{schwarz2010safari}. Interestingly, it was observed that increasing the number of trees in the forest does not lead to overfitting \cite{biau2016random}, just to a more accurate estimation. }
The RF regressor was already applied in several use cases with promising results, e.g. 
\cite{el2022random, huang2020travel,cootes2012robust,nakashima2019passenger}. For our purpose we train the RF on the whole data cube. 
%For our special case, it is of great advantage that the method does not train its trees on the entire data available. As we aim to create an estimator for a data cube cell to detect outliers, there is a great threat of overfitting.
%Overfitting would imply learning the outliers as well as the inliers and thus, being not able to detect anomalous data. Because of its structure, the RF has a smoothing effect on its predictions and therefore seems to be a highly suitable choice as estimator. 
We propose the RF-based estimator as an instantiation of  $\hat{y}\textsubscript{RF}$ and based on this we implemented our RODD-RF detection algorithm. 

A downside of $\hat{y}\textsubscript{RF}$ is that it is computationally more intensive than mean-based estimators.
We therefore also consider other less computationally intensive estimators as our main competitors. As mentioned before, the trimmed mean is very intuitive. Moreover, it is easy to compute and popular in robust statistics \cite{welsh1987trimmed}. 
The SelfExp method works with a trimming percentage of $12.5\%$ and thus uses 75\% of the data %, it does not take into account
%As shown in equation \ref{equation: trimmed mean}, the SelfExp method logarithmizes the values of the data cells and calculated trimmed means. The trimming percentage was originally 75\%, thus the lowest and highest 12.5\% of the values 
%were not taken into account 
for the mean calculation. The choice for this trimming percentage was not justified by \cite{sarawagi1998discovery} and other choices might be at least equally plausible. For example, the trimmed mean that uses 90\% of the data was shown to achieve good results in several studies \cite{hill1982robustness, rocke1982robust,spjotvoll1980comparison}. We thus propose $\hat{y}\textsubscript{S90}$, a trimmed mean estimator with 5\% trimming percentage. %which has been shown to be reliable. 
%This estimator cuts off 15\% less observations compared to the original SelfExp approach.
%To also evaluate the effect of larger trimming percentages, 
We additionally evaluate a trimmed mean estimator $\hat{y}\textsubscript{S60}$ with $20\%$ trimming percentage.
%that thus includes only 60\% of the data. We call this estimator $\hat{y}\textsubscript{S60}$.
Finally, we also consider the median as robust location measure $\hat{y}\textsubscript{Median}$.

%Thus, other trimming percentages were considered as alternatives. We added and subtracted each 15\% to the original 75\%, resulting in trimming percentages of 60\% and 90\%. The corresponding estimators are $\hat{y}\textsubscript{S60}$
%and $\hat{y}\textsubscript{S90}$. 
%The authors of \cite{leys2013detecting} promote the usage of the Median Absolute Deviation method instead of the standard deviation for outlier detection. They showed it is a more robust dispersion measure. Although it could not be used due to a possible loss of comparability, the median could be used as estimator $\hat{y}\textsubscript{Median}$. Subsequently, the standard deviation measured the deviations from the median and not the trimmed mean. 

\section{Simulation Study -- Experimental Setup}
To evaluate the performance of all the presented RODD instantiations, 
we conducted a simulation study. We used synthesized data since using a real-word data cube is not possible because of the missing ground truth of outliers.
%Although we had access to a real-world data cube, we did not use it for performance evaluation because we had no knowledge about the ground truth of outliers.
%Thus, we synthesized data for our simulation study.
The study focused on comparing the quality of the estimators 
 $\hat{y}\textsubscript{S75}$, 
 $\hat{y}\textsubscript{RF}$,
 $\hat{y}\textsubscript{S90}$,
$\hat{y}\textsubscript{S60}$ and
$\hat{y}\textsubscript{Median}$
 for predicting the value of a data cube cell.
 We calculated the classification metrics sensitivity, specificity and accuracy. 
%The importance of the different metrics varies in use cases. In medical applications, the type II error, which produces false negative classifications, may have grave consequences. An example of a type II error would be diagnosing an ill patient as healthy and thus failing the opportunity to treat the illness. The sensitivity is a metric that decreases with an increasing number of false positives and is therefore an important measure in medicine. 
%In medical applications the sensitivity is of great importance \cite{8364558} because not discovering an illnesses and failing the opportunity to treat it leads to grave consequences.
%Specificity plays a more crucial role in credit risk assessment. For fairness reasons, it is important not to deny someone a credit who will likely pay it back. This would be a type I error, a false positive classification.
%There is always a trade-off between sensitivity and specificity. The more sensitive a classification is, the more likely it is to produce false positive classifications and thus, decrease the specificity.
%Accuracy is a suitable metric when there is no need for emphasizing one classification error over the other.
However, the disadvantage of these three evaluation metrics is that they are dependent on a threshold value, which in this case was set to $2.5$. 
To avoid dependency on this threshold value, we also measured the area under the ROC curve (AUC). The ROC curve plots the trade-off between sensitivity and (1 - specificity) varying the threshold parameter $\tau$. If the AUC score is 1, the classification is perfect, if the score is 0.5, the classification is not better than random guessing.

Moreover, we used an ANOVA  for statistical comparison of the outlier detection methods \cite{searle2016linear}.
Their effect and the effect of the further experimental parameters on the AUC score was tested. 
An ANOVA is a statistical method to determine if the means of groups differ significantly from each other \cite{searle2016linear}. It is similar to the t-test, only that it is capable of comparing more than two groups. 

As there are, to our knowledge, no publicly available data cubes that contain a ground truth of outliers, we have synthesized data cubes. In the following, the synthesization is be explained and for better understanding, an example is be provided. 
Eight data cubes, each with three categorical dimensions, were synthesized. The number of dimensions was chosen according to a common application example of outlier detection in sales data (see Section 3). The amounts of different values per dimension were chosen based on a real-world data set from a German household devices selling company. 
Table \ref{tab:art_ex} displays step-by-step how the data cube cells were constructed. 
First, for every value of each dimension an expected value was selected from a random range of numbers (see Table \ref{tab:art_ex_01}). The range varied for each dimension and for each data cube. In the example, the expected value of monthly sales of the product vacuum cleaner is 98, of the city Osaka 110 and for the month of January 91. 
Due to interaction effects, which can be found in real examples of sales figures, expected values for combinations of values of two different dimensions were synthesized (see Table \ref{tab:art_ex_02}). The arithmetic mean of the expected values and the interaction effects was calculated and rounded (see Table \ref{tab:art_ex_03}). For some combinations, there is an interaction effect, but for others, there is not, as the noise was rounded and often resulted in zero.
For the calculation of the expected value of a vacuum cleaner in Osaka, the expected values of the dimensions, 98 and 110 were added. Then, the interaction effect of 4 was added and this sum was divided by 2. The rounded result was 106. 
The expected value for the sales of vacuum cleaners in January is simply the rounded mean of 98 and 91 because there is no interaction effect between this specific product and the month.
To calculate the expected value of sales in Osaka in January 110, 91 and 4 were added and then divided by 2, resulting in 103. 
Finally, the value for one cell was calculated and rounded considering the expected values of each category and the interaction effects (see Table \ref{tab:art_ex_03}). As this value perfectly matched the expected value for the cell, it can be described as noiseless. Due to the calculation process, the expected values approximately follow a normal distribution.

\begin{table}
\begin{subtable}[h]{0.375\textwidth}
\center
\begin{tabular}{ |c||c||c| }
 \hline
 Dimension & Category & Value\\
 \hline
Product & VC & 98\\
Location & Osaka & 110 \\
Location & Berlin & 87\\
Time & January & 91 \\
Time & Febuary & 93\\
 \hline
\end{tabular}
\caption{Random selection of expected values for each dimension.}
\label{tab:art_ex_01}
\end{subtable}
\hfill
\begin{subtable}[h]{0.625\textwidth}
\center
\begin{tabular}{ |c||c||c|  }
 \hline
 Combination & Interaction & Value \\
 \hline
VC $\cap$ Osaka & 4&106\\
VC $\cap$ Berlin & 0 &93\\
VC $\cap$ January & 0&95\\
VC $\cap$ February & -3&94\\
Osaka $\cap$ January & 4&103\\
Osaka $\cap$ February & 0&102\\
Berlin $\cap$ January & 0&89\\
Berlin $\cap$ February & 5&93\\
 \hline
\end{tabular}
\caption{Random selection of interaction effects.}
\label{tab:art_ex_02}
\end{subtable}

\begin{subtable}[h]{\textwidth}
\begin{tabular}{ |c||c||c||c||c| }
 \hline
 Product & Location & Time & Calculation & Value\\
 \hline
VC & Osaka & January & (98 + 110 + 91 + 106 + 95 + 103) / 6 & 101 \\
VC & Osaka & February & (98 + 110 + 93 + 106 + 94 + 102) / 6 & 101\\
VC & Berlin & January & (98 + 87 + 91 + 93 + 95 + 89) / 6 & 92 \\
VC & Berlin & February & (98 + 87 + 93 + 93 + 94 + 93) / 6 & 93\\
 \hline
\end{tabular}
\caption{Xalculation of the noiseless data cube,  cell values are rounded.}
\label{tab:art_ex_03}
\end{subtable}

\caption{Example for the construction of an artificial, noiseless data cube.}
\label{tab:art_ex}
\end{table}

% \vspace{-0.8cm}
For the creation of outliers, a small sample was taken from the data cubes. Based on the remaining cells, the arithmetic mean and the interquartile range were calculated separately for each value of each categorical variable. 1.5 times the interquartile range was subtracted and added to the respective arithmetic mean, resulting in two ‘outlier boundaries’  per value of a dimension. %These boundaries were used for turning the small sample, into outliers. 
For each cell, the rounded outlier boundary that was closest to the original noiseless cell value, was selected and replaced the original value. Subsequently, the values in the sample would be classified as outliers according to the Interquartile Range technique, when looking at each dimension separately. In order to make the data more realistic, integers were added as noise to both outliers and inliers. The noise was calculated as the product of a random number and a fraction $\in \{0.25\%, 1\% or 5\%\}$ of the standard deviation of all values in the data cube. %The random number varied for each cell in the data cube. In each of the eight data cubes, the percentage of outliers was 0.25\%, 1\% or 5\%. 
In order to consider the distribution of the sales values when adding noise, the standard deviation was calculated for each data cube. Then, the standard deviation was divided by 2.5, 5, 7.5, 10 and 12.5, respectively, afterwards it was multiplied with a random integer between -10 and 10. 
For the split of the data cubes into outliers and inliers 30 different random seeds were used. This resulted in \num{3600} data cubes, on which the RODD methods were tested on. 

For our RODD-RF method, we used the implementation of the RF regressor from the Python library sklearn was used \cite{scikit-learn}.
For computational reasons, the hyper-parameter tuning of the RF was performed on a small subset of 120 datacubes using 50 Random Search on 6 parameters (see below). The resulting best parameters were used for all other RFs. This is realistic since performing a new parameter tuning before each application of RODD-RF would be too time-consuming. 
The tuned RF consisted of \num{1500} trees. The number of features was limited to the square root of the total number of features in the data set. The maximal depth of the RF was set to 60, the minimal sample split to 5. The minimum number of samples required to be at a leaf node was 1. For the other parameters, the default values were used. When building the trees bootstrap samples were used. 
This led to a large forest consisting of many deep trees. Each (deep) tree tends overfit the sample it was trained on a lot. In this way, many patterns within the data were captured by a tree. Only if the sample contained outliers, a distortion in the estimation of the cell value could be caused. %However, this did not pose a problem due to the extremely high number of trees.%: As the RF provides an estimator by averaging over all estimators from the trees. Thus, it is quite robust against outliers.
However, this overfitting is later corrected by averaging over the large amount of trees, leading to a good detection of the outliers.

\section{Simulatin Study -- Results}
The experiments showed that there was no single superior method, but for every metric, another RODD method was best for outlier detection (see Table \ref{tab:Metrics_O_M}). The highest sensitivity was achieved using $\hat{y}\textsubscript{S60}$. 
A nearly perfect specificity was the result of using $\hat{y}\textsubscript{Median}$.
The accuracy was equally high for both the $\hat{y}\textsubscript{S75}$ and $\hat{y}\textsubscript{S90}$ approaches. The AUC score was best when applying RODD-RF. Differences between $\hat{y}\textsubscript{S60}$, $\hat{y}\textsubscript{S75}$ and $\hat{y}\textsubscript{S90}$ were generally small. 
\begin{table}
\center
\begin{tabular}{ |p{1.9cm}||p{1.9cm}|p{1.9cm}|p{1.9cm}|p{1.9cm}|  }
 \hline
 Estimator& Sensitivity&Specificity& Accuracy&AUC Score\\
 \hline
 $\hat{y}\textsubscript{S75}$   &0.2079& 0.9900 &\textbf{0.9817}&0.6985 \\
 $\hat{y}\textsubscript{S60}$ & \textbf{0.2116} & 0.9984 & 0.9813& 0.6998\\
$\hat{y}\textsubscript{S90}$ & 0.1899 & 0.9993 & \textbf{0.9817} & 0.6947 \\
$\hat{y}\textsubscript{Median}$ & 0.0032 & \textbf{0.9998} & 0.9790 & 0.5848 \\
$\hat{y}\textsubscript{RF}$ & 0.2012 & 0.9951 & 0.9778 & \textbf{0.7222} \\
 \hline
\end{tabular}
\caption{Evaluation metrics for the five different RODD methods. 
%for outlier detection
}
\label{tab:Metrics_O_M}
\end{table}
%\vspace{-0.8cm}

An factorial ANOVA was run to evaluate the influence of the experimental parameters.  Due to the high number of simulations, all p-values were numerically almost zero.%, i.e. all parameters highly significant.
The $F$ values showed that the chosen estimator $\hat{y}$, the basis dataset, the noise and the percentage of outliers had an impact on the outlier detection performance.  Following up on the result of the ANOVA, t-tests were performed for each parameter. The results of the t-tests on the effect of the estimator on the AUC score and the level of noise are displayed in Table \ref{tab:t_tests_all} (we omit the test results for the basis data set here). 
Both $\hat{y}\textsubscript{S60}$ and $\hat{y}\textsubscript{RF}$ achieved a higher AUC score than $\hat{y}\textsubscript{S75}$, which served as reference model. 
However, $\hat{y}\textsubscript{RF}$ is the estimator with the largest positive effect. Moreover, its results even differ significantly from those of $\hat{y}\textsubscript{S75}$ at an error level of  $0.1\%$ . The p-value of the $\hat{y}\textsubscript{S60}$ is $0.2376$ and thus, it is not significantly different from $\hat{y}\textsubscript{S75}$.

%The t-test on the outlier detection methods (see table \ref{tab:t-test_methods}) showed that both SelfExp60 and the RF estimator achieved a higher AUC score than the original SelfExp method, which served as reference model. 

%Data set 1 is the reference for the other data sets and it is striking that the AUC score is on overage higher in all other datasets. This might be due to the fact that data set 1 is relatively small, thus the methods have not many data to learn expected values from.

For the t-test on the impact of noise on the AUC score, the subset with very much noise served as the reference. Naturally, the AUC score increases with decreasing amount of noise. For the amount of outliers in the data set, the subset with 0.25\% serves as reference. As expected, the more outlier are present the more complicated it is to find them, i.e., the AUC score decreases. % The more noise the data contained, the harder it is to differentiate between inliers and outliers. 

\begin{figure}
    \centering
    \includegraphics[scale=0.35]{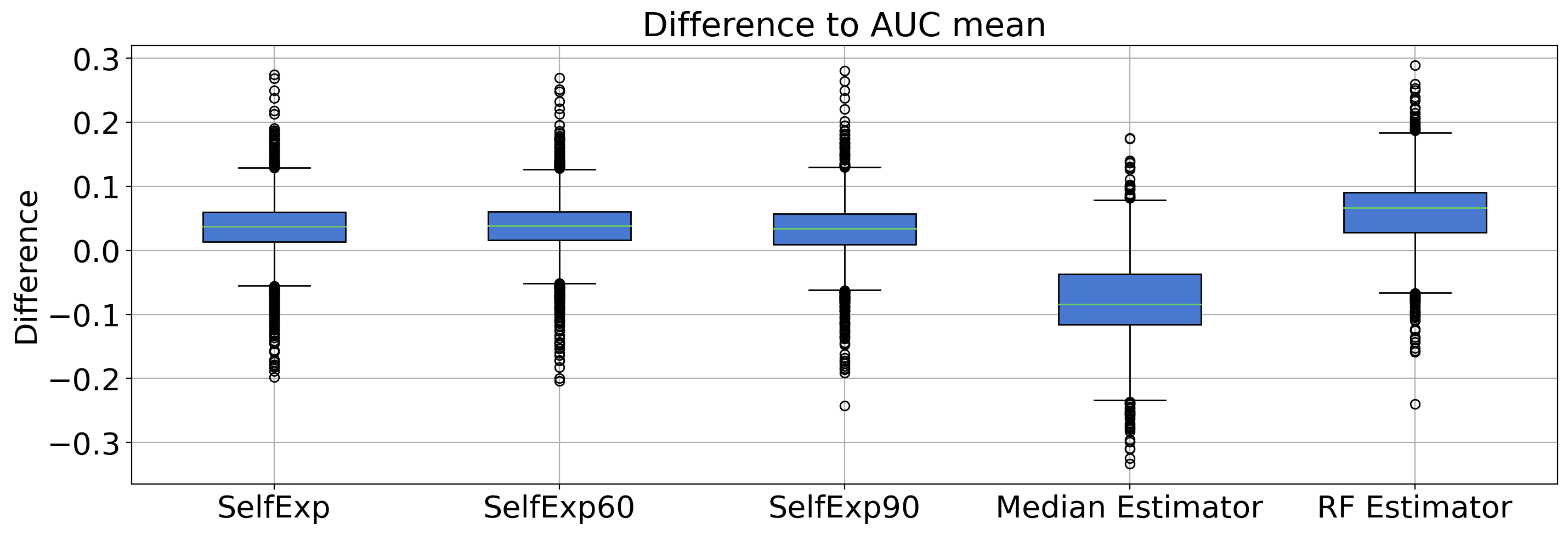}
    \caption{Boxplot comparing the AUC scores of the five different RODD methods.}
    \label{fig:Boxplot_methods}
\end{figure}

\begin{table}
\begin{subtable}[h]{0.32\textwidth}
\center
\begin{tabular}{ |c||c| }
  \hline
 Estimator& Effect \\
 \hline
 $\hat{y}\textsubscript{S60}$ & $0.0014$\\
$\hat{y}\textsubscript{S90}$ & $-0.0038^{**}$\\
$\hat{y}\textsubscript{Median}$ & $-0.1136^{***}$ \\
$\hat{y}\textsubscript{RF}$ & $0.0256^{***}$\\
 \hline
\end{tabular}
%\caption{t-test on the effect of the outlier detection method on the AUC score, with the original method as reference}
\label{tab:t-test_methods}
\end{subtable}
%\hfill
%\begin{subtable}[h]{0.5\textwidth}
%\center
%\begin{tabular}{ |c||c|  }
%hline
% Dataset& Effect\\
% \hline
% 2 & 0.0284^{***} \\
% 3 & 0.0036^{**}\\
% 4 & 0.0135^{***}\\
% 5 &  0.0099^{***}\\
% 6 & 0.0231^{***}\\
% 7 & 0.0175^{***}\\
% 8 & 0.0293^{***}\\
%\hline
%\end{tabular}
%\caption{t-test on basis datasets}
%\label{tab:t-test datasets}
%\end{subtable}
\begin{subtable}[h]{0.32\textwidth}
\center
\begin{tabular}{ |c||c| }
 \hline
 Noise& Effect\\
 \hline
much noise & $0.0598^{***}$\\
moderate noise &  $0.1476^{***}$\\
little noise & $0.2379^{***}$\\
very little noise & $0.3033^{***}$\\
 \hline
\end{tabular}
%\caption{Deterministic calculation of the noiseless data cube, all cell values are rounded to the next integer.}
\label{tab:t-test noise}
\end{subtable}
\begin{subtable}[h]{0.32\textwidth}
\center
\begin{tabular}{ |c||c| }
 \hline
 Outlier & Effect\\
 Percentage & \\
 \hline
 1\%& -0.0011\\
 5\%& $-0.0206^{***}$ \\
 \hline
\end{tabular}
%\caption{Deterministic calculation of the noiseless data cube, all cell values are rounded to the next integer.}
%\label{tab:t-test noise}
\end{subtable}
%\caption{T-tests assessing the impact of the experimental parameters on the AUC score, *** mark a significant effect at the level of 0.1\% and ** mark a significant effect at the level of 1\%.}
\caption{t-tests on the effect of the chosen RODD method on the AUC (left) the level of noise (middle) and the percentage of outliers 
%in the data 
(right). *** mark a significant effect at the level of 0.1\% and ** 
%mark a significant effect 
at the level of 1\%. }
\label{tab:t_tests_all}
\end{table}
%\vspace{-0.8cm}
In order to visualize the different AUC scores, they were adapted. As shown by the ANOVA and the t-tests, the experimental parameters have an impact on the AUC score and thus, might deter the results. For every parameter combination, the mean AUC over all methods and random seeds was calculated. Then, for every individual AUC score, the corresponding mean was subtracted. The resulting comparison is illustrated in Figure \ref{fig:Boxplot_methods}. 
The boxplots visualize the similarity of the estimators $\hat{y}\textsubscript{S75}$, $\hat{y}\textsubscript{S60}$ and $\hat{y}\textsubscript{S90}$
and also show the superiority of RODD-RF ($\hat{y}\textsubscript{RF}$). The estimator $\hat{y}\textsubscript{Median}$ achieved a notably lower AUC score. 

Beside the performance of RODD methods, their run time was also evaluated. 
% In Table \ref{tab:Run_time} the minimal run time for the smallest data cube with \num{3000} cells, the average run time and the maximal run time required for the data cubes with \num{52000} cells are displayed.
The $\hat{y}\textsubscript{S75}$, $\hat{y}\textsubscript{S60}$, $\hat{y}\textsubscript{S90}$ and $\hat{y}\textsubscript{Median}$ estimators had very similar run times. Their minimal run time was 0.13 seconds and on average they took between 0.39 and 0.43 seconds. The maximal run time was between 1.00 and 1.49 seconds.   
%The run time for $\hat{y}\textsubscript{Median}$ is slightly higher and 
The calculation of $\hat{y}\textsubscript{RF}$ required considerably more time. The run time was at minimum 4.08 seconds, on average 37.92 seconds and at maximum \num{1592.05} seconds. 
RFs always require a non-neglectable amount of run time, but the hyper-parameter of the RF models used in the simulation study contribute immensely to increasing the coputation time. For example, the default value for the amount of trees in a forest is 100, but based on the hyper-parameter tuning, we set it to \num{1500}. However, without the parameter tuning the performance of the RF was considerably lower.
Real-world data cubes might be substantially larger than the test data cubes. In this case, it is advisable to evaluate if the increase in the AUC score is worth the increase in computational effort. 
A run time, which can rise up to \num{1592} seconds ($26$ minutes), might make the method unusable, especially if the application occurs 
%of outlier detection methods on data cubes 
frequently.

\section{Application to Real Data}

We also validated the RODD method on a real-world dataset provided by a German household-device selling company. The first five columns of 
Table~\ref{tab:Real-world_ex} give an overview of the provided information on a
subset of the data. 

\begin{table}
\center
\begin{tabular}{ |p{0.8cm}|p{1.5cm}|p{1.4cm}|p{1.8cm}|p{1.2cm}|p{1.4cm}|p{1.2cm}|p{1.4cm}|  }
 \hline
 Year&Month&Product&Distribution Channel&Actual Sales&Expected Sales&SelfExp&Deviation\\
 \hline
2017&January&product3&channel1&5182&3917&2.8&+32.3\%\\
2017&February&product3&channel1&4288&2547&3.9&+68.4\%\\
2017&March&product3&channel1&5056&3675&3.1&+37.6\%\\
2017&April&product2&channel1&4099&2822&2.8&+45.3\%\\
2019&November&product1&channel1&3287&5069&3.9&-35.2\%\\
2019&November&product1&channel2&4883&2379&5.5&+105.3\%\\
 \hline
\end{tabular}
\caption{Subset of the household-device sales data cube together with expected sales and SelfExp values by the RODD-RF method.}
\label{tab:Real-world_ex}
\end{table}
%\vspace{-0.8cm}
The data contained information about the sales numbers of three different products, sold via three different distribution channels (e.g. local store and online shop). The sales from the years 2017, 2018 and 2019 were aggregated on a monthly basis, resulting in 324 datapoints. As the company itself had no information available about outliers in their data, no ground truth could be assumed and we could only do a plausibility check on the identified outliers. The RODD-RF method found 17 outliers, 12 times the sales were higher than expected and 5 times lower. When investigating the outliers, experts from the company found plausible explanations for each of them. It happened twice that a marketing campaign shifted the sales from one channel to another, leading to exceptionally high sales of a product in one channel, while decreasing the sales of the same product in another channel.  In the first case, the sales increased by 105\% in channel 1 compared to the expected value while it was reduced by 35\% in channel 2 (see rows 5 and 6 of Table \ref{tab:Real-world_ex}). Moreover, a direct discount on a product and the reduction of the price of a set proved to be very effective marketing measures. The direct discount increased the sales by 45\% (see row 4 of Table \ref{tab:Real-world_ex}) and the set offer by between 32\% and 68\% (see rows 1-3 of Table \ref{tab:Real-world_ex}).

%, e.g. when checking for changes in the data generation process of live data streams, this may be the case. 
% \begin{table}
% \center
% \begin{tabular}{|p{2.5cm}||p{2.5cm}||p{2.5cm}||p{2.5cm}|}
%  \hline
%  Estimator & Min. run time & Average run time & Max. run time\\
%  \hline
%  $\hat{y}\textsubscript{S75}$   &0.13 seconds & 0.39 seconds & 1.01 seconds \\
%  $\hat{y}\textsubscript{S60}$ & 0.13 seconds & 0.43 seconds & 1.49 seconds\\
% $\hat{y}\textsubscript{S90}$ &0.13 seconds & 0.39 seconds & 1.00 seconds\\
% $\hat{y}\textsubscript{Median}$ & 0.13 seconds& 0.43 seconds &1.45 seconds\\
% %$\hat{y}\textsubscript{RF}$ & 0.20 seconds & 4.83 seconds & 33.88 seconds\\
% $\hat{y}\textsubscript{RF}$ & 4.08 seconds& 37.92 seconds& \num{1592.05} seconds\\
%  \hline
% \end{tabular}
% \caption{Run time of the outlier detection methods.}
% \label{tab:Run_time}
% \end{table}

\section{Conclusion}

We propose RODD a general framework for robust outlier detection in data cubes. RODD subsumes the current gold-standard, the so-called \textit{SelfExp} method \cite{sarawagi1998discovery}, as special case. As first publication we present a systematic comparison with $4$ new robust estimators for outlier detection. Each differs in the statistical estimation of the cell value. While \cite{sarawagi1998discovery} is using a trimmed mean with a $12.5\%$ trimming percentage ($\hat{y}\textsubscript{S75}$) we analyze approaches based on $5\%$ ($\hat{y}\textsubscript{S90}$) and $20\%$ trimming ($\hat{y}\textsubscript{S60}$), the median ($\hat{y}\textsubscript{Median}$) as well as a Random-Forest ($\hat{y}\textsubscript{RF}$, RF-RODD). 
%We evaluated outlier detection methods in datacubes, which has, to the best of our knowledge, not been done previously.
We compared all five methods in a simulation study on \num{3600} data cubes.
%compared the AUC score, the sensitivity, specificity, accuracy as well as the run time of the 
Using AUC score as quality measure %, which has the advantage of being independent of an outlier threshold,
 we observed best results with our new RODD-RF.
 %the RF based estimator $\hat{y}\textsubscript{RF}$. %It was again proven that the RF regressor is a reliable estimator. 
 Compared to traditional \textit{SelfExp} \cite{sarawagi1998discovery}, this effect was highly significant ($p$ value $<0.1\%$). For the other methods, the positive effect on the AUC was not so pronounced ($\hat{y}\textsubscript{S60}$) or even negative ($\hat{y}\textsubscript{90}$, $\hat{y}\textsubscript{Median}$). 
 
 Taking these results into account, we apply ROOD-RF to a real world data set. No labels for the anomalies are available for this data set. However, it turns out that for all anomalies found by the algorithm, an explanation could also be found by human expert and domain knowledge.
% Although it has a considerably longer run time than the other estimators, it is advisable to use it. Also for the other metrics, one of the alternatives always performed better than the original $\hat{y}\textsubscript{S75}$ estimator.

%Moreover, different outlier detection methods were assessed in a simulation study and it was shown which method achieves the best results for which evaluation metric. 

For future work, we plan to incorporate higher dimensional data cubes with more than three categorical dimensions. %While data cubes are a very restrictive environment, they provide also 
Moreover, another measure for the standardization of the difference betweeen the actual and the estimiated value 
%(see Equation~\eqref{eqn:selfexp}) 
could be considered, e.g. a quantile based approach.
Furthermore, we see high potential in hierarchical learners that are less explored in machine learning literature. %We believe that hierarchical learners are a key concept to be explored in future research on robust outlier detection for data cubes.

\vspace{0.5cm}
\textbf{Acknowledgements.} This work was supported by the Research Center Trustworthy Data Science and Security, an institution of the University Alliance Ruhr.

\bibliographystyle{splncs04}
\bibliography{references}




% LNCS Homepage, \url{http://www.springer.com/lncs}. Last accessed 4
% \end{thebibliography}
\end{document}